# Security-Constrained Multi-Objective Optimal Power Flow for a Hybrid AC/VSC-MTDC System with Lasso-based Contingency Filtering

**Yahui Li, Yang Li, Senior Member, IEEE**

School of Electrical Engineering, Northeast Electric Power University, Jilin, China

Corresponding author: Yang Li (e-mail: liyang@neepu.edu.cn).

This work was supported in part by the China Scholarship Council (CSC) under Grant 201608220144, and the National Natural Science Foundation of China under Grant No. 51677023.

**ABSTRACT** In order to coordinate the economy and voltage quality of a meshed AC/VSC-MTDC system, a new corrective security-constrained multi-objective optimal power flow (SC-MOPF) method is presented in this paper. A parallel SC-MOPF model with *N*-1 security constraints is proposed for corrective control actions of the meshed AC/DC system, in which the minimization of the generation cost and voltage deviation are used as objective functions. To solve this model, a novel parallel bi-criterion evolution indicator based evolutionary algorithm (BCE-IBEA) algorithm is developed to seek multiple well-spread Pareto-optimal solutions through the introduction of parallel computation. In this process, a least absolute shrinkage and selection operator (Lasso)-based *N*-1 contingency filtering scheme with a composite security index is developed to efficiently screen out the most severe cases from all contingencies. And thereby, the best compromise solutions reflecting the preferences of different decision makers are automatically determined via an integrated decision making technique. Case studies in the modified IEEE 14- and 300-bus systems demonstrate that the presented approach manages to address this SC-MOPF problem with significantly improved computational efficiency.

## NOMENCLATURE

| | | | |
|---|---|---|---|
| $Y_i$ | admittance between AC grid and converter station $i$ | $R_i$ | slope of converter $i$ |
| $B_{f_i}$ | AC filter of the $i$th converter station | $I_{dc_i}$ | injected current of bus $i$ in the DC grid |
| $\dot{U}_{s_i}$ | voltage phasor of the PCC bus at converter $i$ | $P_{dc_i}$ | injected active power of bus $i$ in the DC grid |
| $\dot{U}_{c_i}$ | voltage phasor of the converter bus at converter $i$ | $I_{dc_{ij}}$ | current flow between bus $i$ and $j$ in the DC grid |
| $U_{dc_i}$ | voltage amplitude of DC bus $i$ | $S_{dc_i}$ | power flow of the $i$th branch in the DC grid |
| $\dot{I}_i$ | current injected into the converter $i$ | $F_0$ | objective function set |
| $S_{s_i}$ | injected apparent power of converter $i$ | $g$ | equality constraint |
| $P_{s_i}$ | active power injected from AC grid to converter $i$ | $h$ | inequality constraint |
| $Q_{s_i}$ | reactive power injected from AC grid to converter $i$ | $x$ | vector of state variables |
| $P_{c_i}$ | active power injected into converter $i$ | $u$ | vector of control variables |
| $Q_{c_i}$ | reactive power injected into converter $i$ | $C$ | considered $N$-1 contingency set |
| $P_{dc_i}$ | active power injected into the DC grid | $C^*$ | critical $N$-1 contingency set |
| $P_{loss,i}$ | loss of the converter $i$ | $f_1$ | function of generation cost |
| $I_{c_i}$ | current magnitude of converter $i$ | $f_2$ | function of voltage deviation index |

| | | | | |
|---|---|---|---|---|
| $N_G$ | number of generators | | $P_{L_i}$ | active power flow of line $i$ |
| $N_{ac}$ | total buses in the AC grid | | $N_C$ | number of reactive power compensation equipment |
| $P_{G_i}$ | active power of generator $i$ | | $N_T$ | number of transformers |
| $U_i$ | voltage amplitude of AC bus $i$; | | $N_{acL}$ | number of AC lines |
| $U_{set,i}$ | preset voltage amplitude of AC bus $i$ | | $N_{dcL}$ | number of DC lines |
| $U_{set,dc_i}$ | preset voltage amplitude of bus $i$ in the DC grid | | $N_{obj}$ | number of objective functions |
| $U_G$ | generator terminal voltage | | $FV(\cdot)$ | fitness value of an individual |
| $T$ | transformer tap-ratio | | $PI_c$ | composite security index |
| $Q_C$ | reactive power compensation capacity | | $A_i$ | alarm limit of the $i$th bus's voltage |
| $P_{g_i}$ | injected active power of bus $i$ | | $H_i$ | security limits of the $i$th bus's bus voltage |
| $Q_{g_i}$ | injected reactive power of bus $i$ | | $P_A$ | upper alarm limit of power flows |
| $P_{d_i}$ | active load of bus $i$ | | $P_H$ | security limit of power flows |
| $Q_{d_i}$ | reactive load of bus $i$ | | $\mathbf{L}$ | response vector |
| $G_{ij}$ | conductance between bus $i$ and $j$. | | $\mathbf{X}$ | matrix consisting of input vectors |
| $B_{ij}$ | susceptance between bus $i$ and $j$ | | $J$ | loss function of fuzzy C-means |
| $\theta_{ij}$ | phase-angle difference between bus $i$ and $j$ | | $N_p$ | numbers of Pareto-optimal solutions |
| $Q_{G_i}$ | reactive power of generator $i$ | | $N_c$ | numbers of clusters |
| $\delta_i$ | voltage angle of bus $i$ | | $d$ | priority membership of grey relational projection |

**Abbreviation**

| | |
|---|---|
| VSC | voltage source converter |
| HVDC | high voltage direct current |
| MTDC | multi-terminal high voltage direct current |
| OPF | optimal power flow |
| PCC | point of common coupling |
| SC-OPF | security constrained optimal power flow |
| CC | corrective control |
| MOPF | multi-objective optimal power flow |
| SC-MOPF | security-constrained multi-objective optimal power flow |
| MOO | multi-objective optimization |
| MOEA | multi-objective evolutionary algorithm |
| BCE | bi-criterion evolution |
| IBEA | indicator based evolutionary algorithm |
| BCE-IBEA | bi-criterion evolution indicator based evolutionary algorithm |
| PBCE-IBEA | parallel bi-criterion evolution indicator based evolutionary algorithm |
| NSGA-II | non-dominated sorting genetic algorithm II |
| MOPSO | multi-objective particle swarm optimization |
| Lasso | least absolute shrinkage and selection operator |
| RPC | reactive power compensation |
| FCM | fuzzy C-means |
| GRP | grey relational projection |
| BCS | best compromise solutions |
| PC | Pareto criterion |
| NPC | non-Pareto criterion |

## I. INTRODUCTION

With increasingly serious energy and environmental problems, it has become a global consensus to promote and advance the transition away from current fossil fuels-based energy pattern to clean, renewable energy sources coupled with greater energy efficiency [1]. For implementing this transition, as an emerging and powerful transmission technology, the voltage source converter (VSC) based high voltage direct current (HVDC) (VSC-HVDC for short) has attracted ever-growing attention since the 1990s and the amount of VSC-HVDC projects dramatically increases in recent years [2]. Compared with conventional current source converter based HVDC, the VSC-HVDC has some significant advantages, such as independent control of active and reactive powers, and controlled islanding [3]. Most importantly, VSC-HVDC offers good prospects for construction of multi-terminal HVDC (MTDC), which makes it suitable for the integration of high-penetration renewable energy sources into smart grids.

### A. BACKGROUND AND MOTIVATION

VSC-MTDC network is capable of providing a cost-effective solution for optimizing the operation of an AC/VSC-MTDC system due to its powerful controllability of power flows. Optimal power flow (OPF) proposed in the 1960s is a classical issue in power systems [4-7], but traditionally it only considers normal operating limits as the constraints. How to ensure the secure operation of power systems has become even more challenging in recent years due to the growing uncertainty resulting from the large-scale integration of new components, such as high-penetration of renewable generations [8] and electric vehicles [9, 10]. As an extension of OPF, security constrained-OPF (SC-OPF) has been considered as a significant tool to balance economy and security of power systems [11], which aims to achieve the economic operation by adjusting the available control variables while stratifying not only normal operating limits, but also violations that would occur during contingencies [12]. In general, SCOPF can be divided into preventive and corrective types, in which the former corresponds to a preventive control action, and the latter is designed for corrective control (CC) [11]. Although the preventive control can enable the system to prevent unplanned operation conditions from occurring, this action normally incurs higher costs due to its inherent conservativeness [13]. Meanwhile, CC has been considered as an effective means for alleviating post-contingency system violations with lower costs [14]. As a result, it is a preferable choice to use CC as a possible control action [11]. On the other hand, mono-objective OPF is becoming unable to meet the diverse needs of optimal operation of power systems. In this context, multi-objective OPF (MOPF) has attracted growing concerns [15-17], since it can coordinate multiple and possibly conflicting objectives. Therefore, this work focuses on the corrective security-constrained MOPF (SC-MOPF) for the AC/VSC-MTDC system.

### B. RELATED WORK

To make full use of VSC-MTDC networks to optimize the power flow the AC/DC system, lots of studies have been performed related to the modeling and optimal operation.

1) *System modeling*: Modeling work of VSC-MTDC was originated in [18], which presented two mathematical models of VSC-MTDC. In [19], a generalized steady state VSC-MTDC model was proposed by Beerten et al. for solving a sequential AC/DC power flow. An open source software for calculating the power flow was developed by the same authors in [20].

2) *Optimal Power Flow*: There have been extensive studies on the OPF problems in VSC-type AC/DC systems [21-24]. The OPF problem was formulated to minimize the transmission loss based on VSC-HVDC system in reference [21]. In [22], the OPF of AC/VSC-HVDC grids was addressed by using the second-order cone programming. Reference [23] has utilized an extended OPF model incorporating VSC-MTDC for the operational cost-benefit analysis. In [24], the information gap decision theory was employed to resolve the OPF issue with consideration of wind farm integration. More recently, some important pioneering works have been reported to address SC-OPF issues in the meshed AC/DC system. In [25], both preventive and corrective SC-OPF have been investigated compared with each other, and the results suggested that the latter yields a cheaper economic dispatch than the former. In [26], an improved corrective SC-OPF was proposed by taking into account $N$-1 security constraints for AC/DC grids, in which a hybrid solution approach was developed. In [14], a hierarchical SCOPF model was developed for a hybrid AC/VSC-MTDC system with high wind penetration. A mixed AC-HVDC test system was presented for the evaluation of SC-OPF algorithms in [27]. Unfortunately, the above works were focused on mono-objective OPF for the AC/DC system.

With the rapid development of artificial intelligence, nature-inspired intelligent computation is becoming a powerful tool for solving many complex power system optimization problems, such as distribution networks [28] and microgrid dispatch [29]. Most recently, multi-objective evolutionary algorithms (MOEAs) have been introduced into solving the MOPF issue of the AC/VSC-MTDC system. As a pioneering work, a MOPF model of the meshed system was developed and the non-dominated sorting genetic algorithm II (NSGA-II) was adopted to solve this model in [30]; and then, this approach was further extended as a two-stage MOPF methodology that incorporates decisions analysis into the multi-objective particle swarm optimization (MOPSO)-based optimization process in [31]. However, the security constraints were not taken into account in these works. In [32], A SC-MOPF algorithm using NSGA-II was proposed to minimize the generation cost and power loss of the AC/MTDC system by considering $N$-1 security constraints.

### C. LIMITATIONS AND CONTRIBUTIONS

Although significant studies in the existing literature have been performed on modeling and control of VSC-MTDC networks to optimize the operation of the meshed AC/VSC-MTDC system, there are still some research gaps in this field as follows. (1) Regarding optimized objectives, recent research suggests that voltage quality is of paramount importance for ensuring secure operation of the system since power flow between VSC-MTDC terminals is determined by DC voltage [14], however, until recently, the issue of coordinating economy and voltage quality of the AC/VSC-MTDC system has attracted rather little attention. (2) For solution methodologies, MOEAs such as NSGA-II in [30, 32] can handle complex multi-objective optimization (MOO) issues, however they, as typical heuristics stochastic optimization algorithms, generally require amounts of computational time, which limits their real-world applications to some extent. (3) Contingency filtering is an important but very challenging task due to the inherent high dimensionality in observations [33-35], especially for a hybrid AC/VSC-MTDC system with numerous elements.

The contributions of this work are mainly as follows.

1) In order to accelerate the computational efficiency when using intelligent optimization algorithms to solve OPF issues, a parallel SC-MOPF model is proposed for a hybrid AC/VSC-MTDC system in this paper through the introduction of parallel computation.

2) An integrated decision making technique is adopted in this work for automatically determining the best compromise solutions reflecting the preferences of different decision makers.

3) The proposed Lasso-based contingency filtering strategy with a composite security index manages to efficiently screen out the most severe cases from all contingencies, and thereby reducing computational load during optimization.

4) The presented approach outperforms other commonly-used MOEAs such as NSGA-II and MOPSO with better optimization performance and significantly higher computational efficiency, which will be demonstrated by using the modified IEEE 14- and 300- bus systems.

## D. ORGANIZATIONAL ARRANGEMENTS

The rest is structured as follows. In Section II, the modeling of VSC-MTDC is briefly introduced. Moreover, the AC/DC SC-MOPF model is formulated in Section III. Section IV presents details of the proposed approach. Case studies are carried out in Section V, and Section VI gives the conclusion.

## II. MODELING OF VSC-MTDC

A simplified model of a meshed AC/VSC-MTDC system with multiple converter stations is shown in Fig. 1 [19].

In this model, $Y_i = G_i + jB_i$ represents the admittance between the AC grid and converter station; $B_{f_i}$ represents the AC filter, which is usually omitted in power flow analysis [19, 20]; $\dot{U}_{s_i} = U_{s_i} \angle \delta_{s_i}$ denotes the voltage phasor of the point of common coupling (PCC) bus at converter $i$; $\dot{U}_{c_i} = U_{c_i} \angle \delta_{c_i}$ is the voltage phasor of the converter bus at converter $i$; and $U_{dc_i}$ is the voltage amplitude of DC bus $i$.

The current injected into the converters is

$$\dot{I}_i = \left(\dot{U}_{s_i} - \dot{U}_{c_i}\right) \cdot Y_i \tag{1}$$

The injected apparent power is given by

$$\dot{S}_{s_i} = P_{s_i} + jQ_{s_i} = \dot{U}_{s_i} \dot{I}_i^* \tag{2}$$

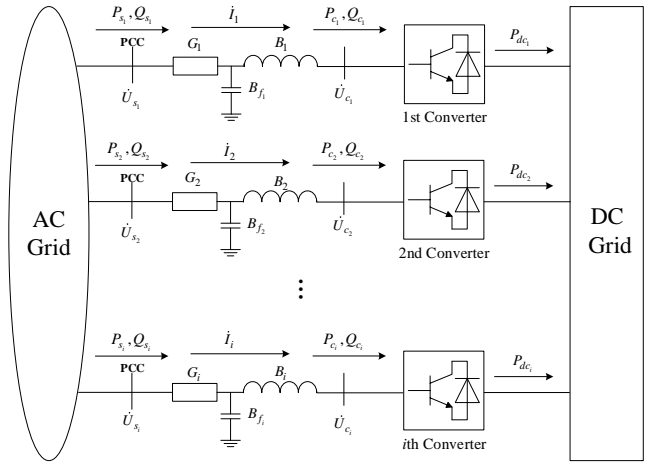

**FIGURE 1.** A simplified model of the meshed AC/VSC-MTDC system.

## A. POWER CHARACTERISTIC OF VSC-MTDC

The active power $P_{s_i}$ and reactive power $Q_{s_i}$ injected from the AC grid are

$$\begin{aligned} P_{s_i} &= U_{s_i}^2 G_i - U_{s_i} U_{c_i} [G_i \cos(\delta_{s_i} - \delta_{c_i}) + B_i \sin(\delta_{s_i} - \delta_{c_i})] \\ Q_{s_i} &= -U_{s_i}^2 B_i - U_{s_i} U_{c_i} [G_i \sin(\delta_{s_i} - \delta_{c_i}) - B_i \cos(\delta_{s_i} - \delta_{c_i})] \end{aligned} \tag{3}$$

Similarly, the active power $P_{c_i}$ and the reactive power $Q_{c_i}$ injected into converter $i$ are

$$\begin{aligned} P_{c_i} &= -U_{c_i}^2 G + U_{s_i} U_{c_i} [G_i \cos(\delta_{s_i} - \delta_{c_i}) + B_i \sin(\delta_{s_i} - \delta_{c_i})] \\ Q_{c_i} &= U_{c_i}^2 B_i - U_{s_i} U_{c_i} [G_i \sin(\delta_{s_i} - \delta_{c_i}) + B_i \cos(\delta_{s_i} - \delta_{c_i})] \end{aligned} \tag{4}$$

The relationship between $P_{c_i}$ and the active power injected into the DC grid $P_{dc_i}$ can be formulated as

$$P_{c_i} - P_{dc_i} - P_{loss,i} = 0 \tag{5}$$

where $P_{loss,i}$ is the loss of converter station $i$, which is

$$P_{loss,i} = a_i + b_i \cdot I_{c_i} + c_i \cdot I_{c_i}^2, \\ I_{c_i} = \sqrt{P_{c_i}^2 + Q_{c_i}^2} \Big/ U_{c_i} \tag{6}$$

where $a_i$, $b_i$ and $c_i$ are the loss coefficients, $I_{c_i}$ denotes current magnitude of converter $i$ [19].

## B. CONVERTER CAPACITY LIMIT

To guarantee safe operation, the operating points of converter stations must be situated within the *PQ*-capability chart [19]. The current and voltage limits of converter station $i$ is

$$r_{i,\min}^2 \leq \left(P_{s_i} - P_{i0}\right)^2 + \left(Q_{s_i} - Q_{i0}\right)^2 \leq r_{i,\max}^2 \tag{7}$$

where $S_{i0}(P_{i0}, Q_{i0})$ is the circle's center, $r_{i,\min}$ and $r_{i,\max}$ are the minimum and maximum limits of the radius $r_i$.

## C. CONTROL OF VSC-MTDC

For a VSC-MTDC terminal, there are several available control modes such as the constant power control, the constant DC voltage control and the droop DC voltage control. In this work, the droop control is chosen since it gives the most satisfactory results in practice [36].

When using this strategy, the slope $R_i$ of converter $i$ needs to be controlled. The slope can be switched to the constant DC voltage or constant power control strategies if it is set to ∞ or 0.

## D. DC GRID MODEL

In the DC grid, the injected current $I_{dc_i}$ of bus $i$ is [19]

$$I_{dc_i} = \sum_{j=1, j \neq i}^{N_{dc}} Y_{dc_{ij}} \cdot (U_{dc_i} - U_{dc_j}) \tag{8}$$

where $N_{dc}$ is the number of DC buses, $Y_{dc_{ij}}$ denotes the admittance between DC buses $i$ and $j$.

The injected active power of bus $i$ is

$$P_{dc_i} = U_{dc_i} I_{dc_i} \tag{9}$$

The currents and voltages obey the following constraints,

$$\begin{aligned} I_{dc_{ij},\min} &\leq I_{dc_{ij}} \leq I_{dc_{ij},\max} \\ U_{dc_i,\min} &\leq U_{dc_i} \leq U_{dc_i,\max} \end{aligned} \tag{10}$$

where $I_{dc_{ij}}$ is the current flow between bus $i$ and $j$, its upper and lower limits are respectively $I_{dc_{ij},\max}$ and $I_{dc_{ij},\min}$; $U_{dc_i,\min}$ and $U_{dc_i,\max}$ are the lower and upper limits of $U_{dc_i}$.

The power flow of a DC line obeys the following constraint:

$$P_{dc_i,\min} \leq P_{dc_i} \leq P_{dc_i,\max}, \quad i=1,\ldots,N_{dcL} \quad (11)$$

where $P_{dc_i}$ is the power flow of line $i$ in the DC grid; $P_{dc_i,\min}$ and $P_{dc_i,\max}$ are respectively the lower and upper limits of $P_{dc_i}$; $N_{dcL}$ is the total number of DC lines.

### III. PROBLEM FORMULATION

The corrective SC-MOPF formulation of AC systems, originally proposed in [37], is here utilized. The used corrective SC-MOPF model that governs the CC actions is

$$\begin{aligned}
\min \quad & F_0(\boldsymbol{x}_0, \boldsymbol{u}_0) \\
s.t. \quad & g_0(\boldsymbol{x}_0, \boldsymbol{u}_0) = 0, \\
& h_0(\boldsymbol{x}_0, \boldsymbol{u}_0) \leq 0, \\
& g_k(\boldsymbol{x}_k, \boldsymbol{u}_k) = 0, \ k \in C \\
& h_k(\boldsymbol{x}_k, \boldsymbol{u}_k) \leq 0, \ k \in C \\
& |\boldsymbol{u}_k - \boldsymbol{u}_0| \leq \Delta \boldsymbol{u}_k^{\max}, \ k \in C \\
& \boldsymbol{u}_0^{\min} \leq \boldsymbol{u}_0 \leq \boldsymbol{u}_0^{\max}.
\end{aligned} \quad (12)$$

where $F_0$ is the objective function set; $g$ and $h$ are the equality and inequality constraints; $\boldsymbol{x}$ is the vector of state variables and $\boldsymbol{u}$ is the vector of control variables; $C = \{1, 2, \cdots, c\}$ represents the $N$-1 contingencies which considers all the outages of AC and DC lines; the subscripts '0' and '$k$' denote pre-contingency and post-contingency states; $\Delta \boldsymbol{u}_k^{\max}$ is the vector of maximally allowed adjustment control variables; $\boldsymbol{u}_0^{\min}$ and $\boldsymbol{u}_0^{\max}$ are the lower and upper limits of the pre-contingency control vector $\boldsymbol{u}_0$.

Strictly speaking, the operating point is correctively secure only if all contingencies are feasible for all of the constraints in (12). However, it is a challenging task to incorporate security constraints in large-scale optimization problems due to their inherently high nonlinearities. For this reason, most of the SC-OPF approaches only consider and examine the severe contingencies, rather than all cases, to reduce the computational complexity. In this work, a critical contingency set $C^*$ is obtained by using contingency filtering [34, 35], and then each contingency in set $C^*$ is checked to determine the feasibility of CC actions. It's worth pointing out that the set $C^*$ is dynamically regulated in term of the change of the operation point.

#### A. OBJECTIVE FUNCTIONS

As a major concern in OPF, the popular generation cost is adopted as an objective function [11, 14, 30-32]. In addition, minimizing voltage deviation is handled as another objective, since for AC/VSC-MTDC systems, it is of the utmost importance to maintain adequate DC voltages during the actual operations [31, 38]. Consequently, the objective function set $F_0$ consists of the generation cost $f_1$ and the voltage deviation index $f_2$, which are

$$\begin{aligned}
\min f_1(\boldsymbol{x},\boldsymbol{u}) &= \sum_{i=1}^{N_G} (\alpha_i P_{G_i}^2 + \beta_i P_{G_i} + \gamma_i) \\
\min f_2(\boldsymbol{x},\boldsymbol{u}) &= \sum_{j=1}^{N_{ac}} (U_j - U_{set,j})^2 + \sum_{k=1}^{N_{dc}} (U_{dc_k} - U_{set,dc_k})^2
\end{aligned} \quad (13)$$

where $N_G$ and $N_{ac}$ are the number of generators and total buses in the AC grid; $P_{G_i}$ denotes the $i$th generator's active power; $\alpha_i$, $\beta_i$ and $\gamma_i$ are the cost coefficients of generator $i$; $U_j$ denotes the $j$th bus's voltage; $U_{set,j}$ and $U_{set,dc_k}$ indicate the preset voltages. The control vector $\boldsymbol{u}$ is formulated as

$$\boldsymbol{u} = [P_G, U_G, T, Q_C, P_s, Q_s, U_{dc}, R] \quad (14)$$

where $U_G$, $T$ and $Q_C$ are respectively the generator voltage, transformer tap-ratio, and reactive power compensation (RPC) capacity. Note that, only $T$ and $Q_C$ are discrete variables, while all other control variables in vector $\boldsymbol{u}$ are continuous variables.

#### B. CONSTRAINTS

This section gives the related constraints in the AC grid.

$$\begin{aligned}
P_{g_i} - P_{d_i} - U_i \sum_{j \in i} U_j \left( G_{ij} \sin \theta_{ij} + B_{ij} \cos \theta_{ij} \right) = 0, i=1,\cdots,N_{ac} \\
Q_{g_i} - Q_{d_i} - U_i \sum_{j \in i} U_j \left( G_{ij} \sin \theta_{ij} - B_{ij} \cos \theta_{ij} \right) = 0, i=1,\cdots,N_{ac}
\end{aligned} \quad (15)$$

where $P_{g_i}$ and $Q_{g_i}$ are the active and reactive power inputs of bus $i$; $P_{d_i}$ and $Q_{d_i}$ are the active and reactive loads of bus $i$; $U_i$ and $U_j$ are respectively the voltage amplitudes of bus $i$ and $j$; $G_{ij}$, $B_{ij}$ and $\theta_{ij}$ are respectively the conductance, susceptance and phase-angle difference between bus $i$ and $j$.

The inequality constraints are

$$\begin{aligned}
P_{G_i,\min} \leq P_{G_i} \leq P_{G_i,\max}, & \quad i=1,\ldots,N_G \\
Q_{G_i,\min} \leq Q_{G_i} \leq Q_{G_i,\max}, & \quad i=1,\ldots,N_G \\
U_{i,\min} \leq U_i \leq U_{i,\max}, & \quad i=1,\ldots,N_{ac} \\
\delta_{i,\min} \leq \delta_i \leq \delta_{i,\max}, & \quad i=1,\ldots,N_{ac} \\
T_{i,\min} \leq T_i \leq T_{i,\max}, & \quad i=1,\ldots,N_T \\
Q_{C_i,\min} \leq Q_{C_i} \leq Q_{C_i,\max}, & \quad i=1,\ldots,N_C \\
P_{L_i,\min} \leq P_{L_i} \leq P_{L_i,\max}, & \quad i=1,\ldots,N_{acL}
\end{aligned} \quad (16)$$

where $Q_{G_i}$ is the reactive power of generator $i$; $\delta_i$ is the voltage angle of bus $i$; $P_{L_i}$ is the active power flow of AC line $i$; the subscripts 'min' and 'max' are the minimum and maximum limits; $N_C$, $N_T$, and $N_{acL}$ are the number of reactive power compensation equipment, transformers, and AC lines.

## IV. SOLUTION METHOD

To solve this SC-MOPF model, a new mixed-coded parallel bi-criterion evolution indicator based evolutionary algorithm (BCE-IBEA), PBCE-IBEA for short, with Lasso-based contingency filtering is proposed for finding the set of Pareto-optimal solutions; and then, an integrated decision making technique combining fuzzy C-means (FCM) clustering with grey relational projection (GRP) is utilized for identifying the best compromise solutions (BCSs) in [31].

### A. PRINCIPLES OF BCE-IBEA

The BCE-IBEA proposed in [39] is based on the bi-criterion evolution (BCE) framework with indicator based evolutionary algorithm (IBEA) embedded into its Non-Pareto criterion (NPC) evolution part. During evolution, the population is guided to evolve fast toward Pareto fronts while maintaining its diversity. It is because this algorithm can utilize the advantages of Pareto criterion (PC) and NPC and compensates for each other's disadvantages, it is chosen to solve the MOPF issue.

The procedures of IBEA and BCE are introduced as follows.

#### 1) INDICATOR BASED EVOLUTIONARY ALGORITHM

As a powerful MOEA, IBEA utilizes a performance indicator to optimize the desired property of the evolutionary population [40]. The main procedures of IBEA are listed as follows.

Step 1: **Initialization**: Initialize the AC/DC system parameters, the population $Pop$ and its size $s$, and assign the current iteration $It$ to 0.

Step 2: **Population formation**: Calculation of the objective function values of all individuals. Calculate the OPF of the AC/DC system via the alternating iterative method proposed in [19], and thereby obtain the values of the objective functions $F = \{f_1, f_2, ..., f_{N_{obj}}\}$, where $N_{obj}$ is the number of the objective functions.

Step 3: **Fitness evaluation**: Different from conventional MOEAs, the fitness evaluation is performed on the basis of a binary additive $\varepsilon$-indicator $I_{\varepsilon^+}$ in the IBEA, which can be utilized for guiding the evolutionary process by measuring the relative approximations of two Pareto sets.

Given two sets $Po_1$ and $Po_2$, $I_{\varepsilon^+}(Po_1, Po_2)$ is defined by [40]

$$I_{\varepsilon^+}(Po_1, Po_2) = \arg\min_{\varepsilon}\{\forall a^2 \in Po_2, \exists a^1 \in Po_1 : \quad (17)$$
$$f_i(a^1) - \varepsilon \leq f_i(a^2), i = 1, ..., N_{obj}\}$$

where $a^1$ and $a^2$ are two individuals, and they respectively belong to $Po_1$ and $Po_2$.

And thereby, the fitness value of the individual $a^1$ is [40]

$$FV(a^1) = \sum_{a^2 \in Pop\setminus\{a^1\}} -e^{-I(\{a^2\},\{a^1\})/\kappa} \quad (18)$$

where $FV(\cdot)$ is the fitness value, and $\kappa$ is a scaling factor.

Step 4: **Environmental selection**: Iterate the following selection until the size of the newly generated population is not greater than $s$: choose an individual in $Pop$ with the smallest fitness value and remove it; update the fitness values of the remaining individuals.

Step 5: **Termination judgment**: If the termination criterion is met, then output the Pareto-optimal solution set, and terminate the process. Here the criterion is whether $It$ exceeds the pre-given maximum iteration number, i.e., $It \geq It_{max}$.

Step 6: **Mating selection**: Execute the binary tournament selection on the population to form the mating pool $Pop'$.

Step 7: **Variation**: The resulting offspring are added to $Pop$ after crossover and mutation operators on $Pop'$. Increase the counter $It$ by 1 ($It = It + 1$) and return to Step 2.

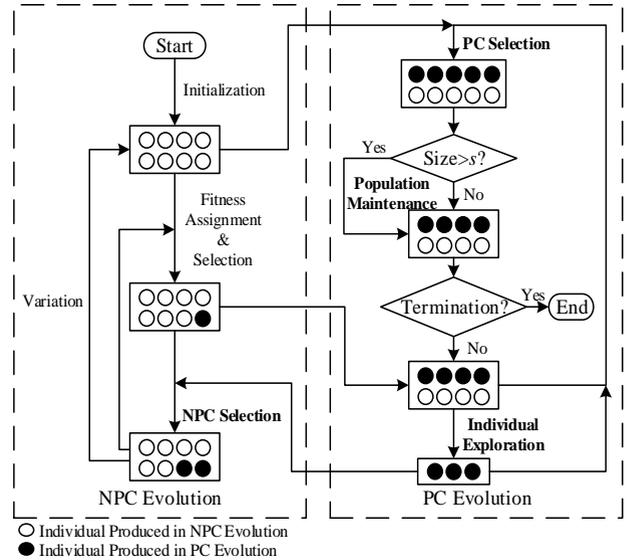

**FIGURE 2.** Flowchart of the BCE.

#### 2) BI-CRITERION EVOLUTION

In BCE, evolutionary populations are divided into the PC and NPC population, which frequently exchange and share information while evolving on the basis of their own criterion. There are four key operations: PC and NPC selection, population maintenance, and individual exploration. The flowchart of the BCE is illustrated in Fig. 2.

As illustrated in Fig. 2, BCE includes the NPC and PC evolution parts. The termination criterion here is whether the iteration number reaches a pre-assigned number of evaluations or not. The more details of BCE-IBEA are described in [39].

### B. CONTINGENCY FILTERING

A Lasso-based contingency filtering approach is proposed to screen out the most severe cases from the contingency list. The composite security index $PI_c$ is used for security assessment of the outage of all AC lines [34], which is defined as

$$PI_c = \left[ \sum_i \left(q_{U_i}^{\max}\right)^{2n} + \sum_i \left(q_{U_i}^{\min}\right)^{2n} + \sum_j \left(q_{P_j}\right)^{2n} \right]^{\frac{1}{2n}}$$

$$s.t. \ q_{U_i}^{\max} = \begin{cases} \dfrac{U_i - H_i^{\max}}{A_i^{\max} - H_i^{\max}}, & if \ U_i > H_i^{\max} \\ 0, & otherwise \end{cases}$$

$$q_{U_i}^{\min} = \begin{cases} \dfrac{H_i^{\min} - U_i}{H_i^{\min} - A_i^{\min}}, & if \ U_i < H_i^{\min} \\ 0, & otherwise \end{cases} \quad (19)$$

$$q_{P_j} = \begin{cases} \dfrac{|P_j| - P_{H_j}}{P_{H_j} - P_{A_j}}, & if \ |P_j| > P_{H_j}; \\ 0, & otherwise \end{cases}$$

where $n$ denotes the exponent ($n=2$); $A_i^{\min}$ and $A_i^{\max}$ are the minimum and maximum alarm limits of the $i$th bus's voltage $U_i$; $H_i^{\min}$ and $H_i^{\max}$ are the minimum and maximum security limits of bus voltage $U_i$; $P_{A_j}$ and $P_{H_j}$ are the upper alarm limit and the security limit of power flow $P_j$ through each line. If $PI_c>1$, the system is insecure; if $0<PI_c\le1$, it is in an alarm state; if $PI_c=0$, it is secure.

The Lasso is based on the following linear model [41]:

$$\mathbf{L} = \mathbf{X}\sigma + \xi \quad (20)$$

where $\mathbf{L} = [L_1,...L_{N_o}]^T$ is the response vector; $\mathbf{X}$ is the $N_o \times N_q$-design matrix which consists of the input vectors $\mathbf{X}_1,...,\mathbf{X}_{N_o}$, and each vector corresponds to one response; $\xi$ is the innovation process modeled as a sequence of random variables. To obtain the $\sigma = [\sigma_1,...,\sigma_{N_q}]$, the following convex optimization problem needs to be solved:

$$\min_\sigma \left( N_o^{-1} \|\mathbf{L} - \mathbf{X}\sigma\|^2 + \lambda \sum_{j=1}^{N_q} |\sigma_j| \right) \quad (21)$$

where $\lambda \geq 0$ is the shrinkage tuning parameter.

In this study, the outage line numbers and the control variables in (14) are used as the inputs, while the corresponding index $PI_c$ is the output. Once the Lasso is trained, it can predict $PI_c$ according to the control vector $\boldsymbol{u}$. The AC lines in an insecure state and all DC lines are selected to constitute the set $C^*$ in this study. By doing so, the most severe cases can be screened out from the entire contingency list.

### C. PARALLEL BCE-IBEA WITH LASSO

In order to accelerate the computation, the PBCE-IBEA is developed by introducing parallel computing technology, where multiple processes perform the following optimization process in a coordinated manner:

Processor 0 is responsible for task assignment and coordination, and executes initialization, selections and variation, and termination judgment.

Processors 1 to $m$-1 execute calculation tasks, where $m$ is the number of processors. In each iteration, once an individual's fitness needs to be calculated, the task is assigned to processors 0 to $m$-1. For each one of these processors, the following operations are executed in parallel:

- Calculate AC/DC power flow and obtain the objective function values;
- Calculate the fitness values of the assigned individuals;
- Screen out the critical contingency set $C^*$ from the entire contingency list by predicting with Lasso;
- Check contingency to determine the feasibility of CC actions. Specifically, calculate power flows considering each post-contingency in $C^*$, and check whether all the constraints in (12) are satisfied based on the results.

By adopting Lasso for contingency filtering, the parallel mechanism with $m$ processors is shown in Fig. 3.

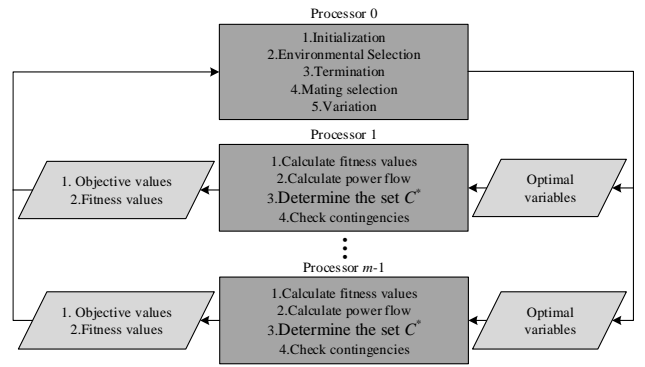

**FIGURE 3.** Parallel computation mechanism of BCE-IBEA.

### D. DECISION ANALYSIS

The obtained Pareto-optimal solutions are divided into different clusters via FCM clustering, and thereby the BCSs in each cluster are determined via the GRP. This decision process is described in more detail in [31].

FCM clustering is modeled as the following issue:

$$\min \ J = \sum_{i=1}^{N_p} \sum_{j=1}^{N_c} \mu_{ij}^m \|w_i - v_j\|^2$$
$$s.t. \quad \sum_{j=1}^{N_c} \mu_{ij} = 1, \ i=1,...,N_p \quad (22)$$

where $J$ denotes the loss function, $N_p$ and $N_c$ are the numbers of the Pareto-optimal solutions and clusters, $\mu_{ij} \in [0,1]$ is the membership degree between solution $w_i$ and clustering center $v_j$, $m \in [1,\infty]$ is a fuzziness control parameter. Here, $N_c$ is taken as 2, which corresponds to the two objective functions.

The GRP method is then utilized to assess the solutions belonging to the same cluster. For a solution $l$, the priority membership $d$ is calculated by [31]

$$d_l = \frac{(V_0 - V_l^-)^2}{(V_0 - V_l^-)^2 + (V_0 - V_l^+)^2}, \quad 0 \le d_l \le 1$$

$$V_l^{+(-)} = \sum_{k=1}^{N_t} \gamma_{lk}^{+(-)} \frac{\omega_k^2}{\sqrt{\sum_{k=1}^{N_t}(\omega_k)^2}} \quad (23)$$

where $V_l^{+(-)}$ is the projection of the *l*th positive (+) or negative (−) ideal reference solution; $V_0$ equals to $V_l$ if $\gamma = 1$; $\gamma_{lk}^{+(-)}$ is grey relational coefficient; $N_t$ is the number of indicators; $\omega_k$ is the weight of *k*th objective function in the solution. To simplify the analysis, the two objectives have the same weight in this study. Note that for two solutions, the membership *d* with a higher value represents a better quality.

## V. CASE STUDIES

To examine the performance of the proposed approach, two test cases have been performed. All programs are developed under the MATLAB environment on a computer with Intel Core i5-4590 3.3 GHz four-core processors and 4 GB RAM.

### A. CASE 1—IEEE 14-BUS SYSTEM

The well-known IEEE 14-bus system in the literature [14, 22, 25, 26, 30, 31] is modified as the test case, which comprises 5 generators, 11 loads, 20 branches (including 17 AC lines, 3 DC lines), 1 RPC (connected to bus 9) and three modular multilevel converters, as shown in Fig. 4.

1) PARAMETER SETTING

The ranges of control variables are set as follows. The bus voltage is within the range 0.90 to 1.10 p.u.; $T$ is in the range from 0.9 to 1.1 with the step 0.0125; the RPC capacity is in the range 0 to 0.5 p.u. with the step 0.01 p.u.; both the $P_s$ and $Q_s$ range from -1.0 to 1.0 p.u.; and $U_{dc}$ is in the range from 0.90 to 1.10 p.u.. The droop slope $R$ in each converter is in the range [-10, 10]. For PBCE-IBEA, the population size and the maximum number of iterations are respectively 100 and 50.

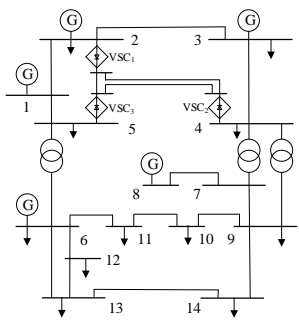

FIGURE 4. Modified IEEE 14-bus system.

2) RESULTS AND DISCUSSION

First, PBCE-IBEA is used to find the Pareto-optimal solutions. One representative set of Pareto-optimal solutions is chosen in 30 independent runs, and its distribution in the objective function space is illustrated in Fig. 5.

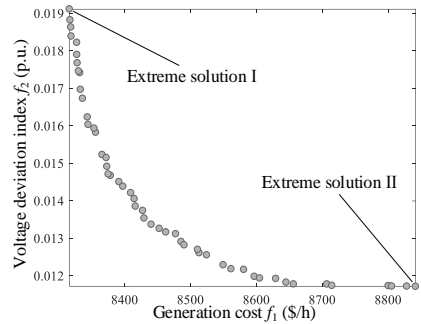

FIGURE 5. Distribution of Pareto-optimal solutions in Case 1.

Fig. 5 suggests that the PBCE-IBEA manages to yield the multiple well-distributed Pareto-optimal solutions. As a result, one can draw a conclusion that the economy and voltage quality can be effectively coordinated by using the proposed method.

And then, FCM clustering is utilized to cluster the representative Pareto optimals into different clusters, as demonstrated in Fig. 6. Note that the points representing $f_1$ and $f_2$ are respectively marked with the red and green color.

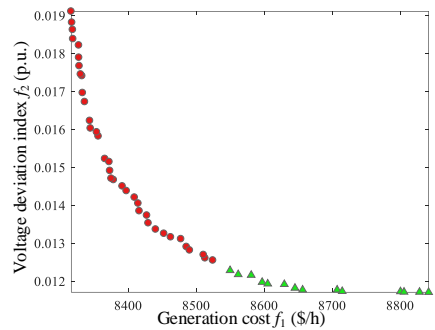

FIGURE 6. Distribution of Pareto-optimal solutions after clustering.

From Fig. 6, it can be observed that the Pareto optimals are separated into two clusters by the FCM clustering.

The GRP is then applied to evaluate the priority memberships $d$, and the solutions with the highest membership in the two groups are chosen as the BCSs, as listed in Table I.

TABLE I
BCSs IN THE IEEE 14- BUS SYSTEM

| BCSs | $f_1$/(\$/h) | $f_2$/(p.u.) | $C^*$ | $d$ |
|---|---|---|---|---|
| BCS I | 8375.49 | 0.0147 | L1(1-2), L4(3-4), DC lines | 0.6872 |
| BCS II | 8656.74 | 0.0118 | L1(1-2), L3(2-3), DC lines | 0.7532 |

In order to examine the efficacy of the Lasso-based contingency filtering scheme, taking the set $C^*$ that corresponds to BCS I for example, all contingencies (except for secure states) are ranked in the order of severity according to indicator $PI_c$ predicted by the Lasso in Table II, where the prediction error $Err$ (%) is calculated by the following formula:

$$Err = (PI_{c,1} - PI_{c,2})/PI_{c,2} \times 100 \quad (24)$$

where $PI_{c,1}$ and $PI_{c,2}$ are the $PI_c$ values, which are respectively obtained by using the Lasso approach and the direct computation according to (19).

TABLE II
COMPARISON OF SCALAR-VALUED COMPOSITE SECURITY INDEX

| Branch number | Security index $PI_c$ | | Prediction error (%) |
|---|---|---|---|
| | $PI_{c,1}$ | $PI_{c,2}$ | |
| L4 | 3.3688 | 3.3386 | 0.9046 |
| L1 | 2.3926 | 2.5000 | -4.2960 |
| L7 | 0.9663 | 0.9843 | -1.8287 |
| L15 | 0.7058 | 0.6749 | 4.5784 |
| L16 | 0.5629 | 0.5589 | 0.7157 |
| L17 | 0.4881 | 0.5004 | -2.4580 |
| L3 | 0.2825 | 0.2857 | -1.1201 |
| L5 | 0.2614 | 0.2723 | -4.0029 |
| L6 | 0.1948 | 0.1858 | 4.8438 |
| L13 | 0.1390 | 0.1452 | -4.2700 |
| L8 | 0.0928 | 0.0907 | 2.3153 |
| L14 | 0.0823 | 0.0855 | -3.7427 |
| L12 | 0.0805 | 0.0828 | -2.7778 |

Table II indicates that the ranking results of contingencies are consistent by using the two methods, and there are no significant differences between the obtained security indexes. Therefore, the Lasso is suitable for $N$-1 contingency filtering.

Taking BCS I as an example, the key variables in the system before and after optimization are shown in Tables III-V.

TABLE III
GENERATOR VARIABLES BEFORE AND AFTER OPTIMIZATION

| Generators | Before optimization | | | After optimization | | |
|---|---|---|---|---|---|---|
| | $P_G$/pu | $Q_G$/pu | $U_G$/pu | $P_G$/pu | $Q_G$/pu | $U_G$/pu |
| G1 | 2.324 | -0.165 | 1.060 | 1.6281 | -0.098 | 1.060 |
| G2 | 0.400 | 0.436 | 1.045 | 0.5759 | 0.252 | 1.045 |
| G3 | 0 | 0.251 | 1.010 | 0.2999 | 0.017 | 1.010 |
| G4 | 0 | 0.127 | 1.070 | 0.0997 | 0.173 | 1.070 |
| G5 | 0 | 0.176 | 1.090 | 0.0971 | 0.232 | 1.090 |

TABLE IV
VARIABLES IN THE DC GRID BEFORE AND AFTER OPTIMIZATION

| VSCs | Before optimization | | | | After optimization | | | |
|---|---|---|---|---|---|---|---|---|
| | $P_s$/pu | $Q_s$/pu | $U_{dc}$/pu | $R$/pu | $P_s$/pu | $Q_s$/pu | $U_{dc}$/pu | $R$/pu |
| $VSC_1$ | -0.8620 | 0.0111 | 1.000 | 0.0050 | -0.7776 | 0.0023 | 1.049 | 0.0023 |
| $VSC_2$ | 0.9680 | -0.1237 | 1.000 | 0.0050 | 0.8712 | -0.1539 | 1.024 | 0.0053 |
| $VSC_3$ | -0.1296 | 0.1353 | 1.000 | 0.0050 | -0.1426 | 0.1429 | 1.067 | 0.0053 |

TABLE V
OBJECTIVE FUNCTION VALUES BEFORE AND AFTER OPTIMIZATION

| Optimization status | $f_1$/($/h) | $f_2$/(p.u.) |
|---|---|---|
| Before optimization | 12602.30 | 0.0204 |
| After optimization | 8375.49 | 0.0147 |

From Tables III-V, it can be observed that the distribution of power flow becomes much better after optimization, which embodies both objectives after optimization are superior to their corresponding values before optimization. Therefore, these results verify our approach's effectiveness on this issue.

### 3) COMPARISON WITH OTHER ALGORITHMS

To properly evaluate the performance of our approach, comparison tests with other popular algorithms, such as the original BCE-IBEA, NSGA-II [30] and MOPSO [31], have been performed. To facilitate comparison, the common parameters of these algorithms, such as the population size, are set in the same way.

Particularly, each algorithm has their specific parameters. In BCE-IBEA and PBCE-IBEA, the specific parameter $\kappa$ is 0.05. The crossover probability and mutation probability, which are the specific parameters of NSGA-II, are respectively 0.9 and $1/L_c$ ($L_c$ is the length of a chromosome). The inertia weight, the learning coefficient, and the divisions for the adaptive grid, which are the specific parameters of MOPSO, are respectively 0.73, 1.5 and 30.

The most representative Pareto fronts in 30 independent runs of each algorithm are shown in Fig. 7.

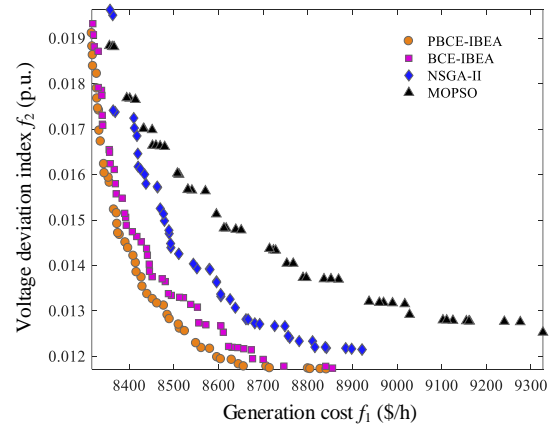

**FIGURE 7.** Pareto fronts of the different algorithms.

Fig. 7 shows that the PBCE-IEBA has the better optimization performance than all comparison algorithms, embodying that its Pareto front dominates the others' fronts in most cases.

To reasonably assess the execution times of these algorithms, 30 independent runs are performed for each algorithm due to the inherent randomness of MOEAs [15], and the obtained average running times are illustrated in Fig. 8.

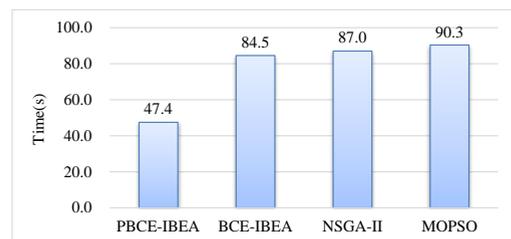

**FIGURE 8.** Running times of different algorithms

Fig. 8 shows that the solution efficiency of the proposed PBCE-IEBA is far superior to that of other alternatives. More specifically, the running time of the PBCE-IEBA is reduced to 56.09%, 54.48% and 52.49% of the original BCE-IEBA, NSGA-II, and MOPSO. It can be expected that it more processors are used, the computational efficiency will be further improved. Therefore, this evidence clearly indicates that parallel computation manages to accelerate the computation.

### B. CASE 2—IEEE 300-BUS SYSTEM

A four-terminal MTDC network is embedded into the modified IEEE 300-bus system [18, 30, 31], which has 69 generators, 68 loads and 411 branches. For ease of presentation, only the MTDC network of this system is illustrated in Fig. 9.

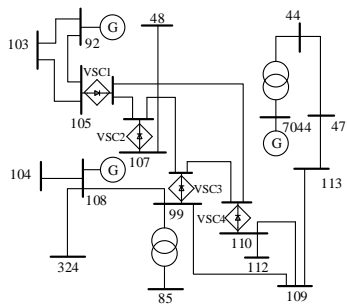

**FIGURE 9.** Relevant part of the modified IEEE 300-bus system.

The ranges of control variables and the algorithm parameters of the PBCE-IBEA in Case 2 are the same as those in Case 1. For solving the SC-MOPF problem, all insure contingencies and DC lines are considered. And the distribution of Pareto-optimal solutions obtained by PBCE-IBEA is shown in Fig. 10.

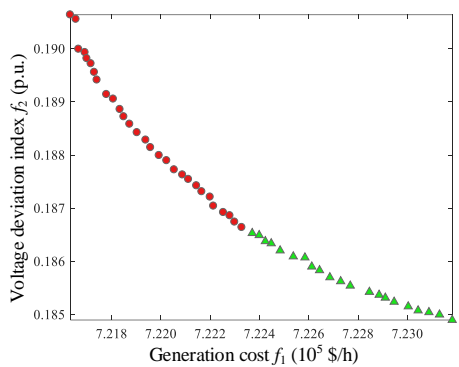

**FIGURE 10.** Distribution of Pareto-optimal solutions in Case 2.

The GRP method is then applied to evaluate the priority memberships of the two groups clustered by FCM, and the solutions with the highest membership are chosen as the BCSs, and the BCSs of this system are shown in Table VI.

TABLE VI
OBTAINED BCSs IN THE IEEE 300- BUS SYSTEM

| BCSs | $f_1$/($/h) | $f_2$/(p.u.) | $d$ |
|---|---|---|---|
| BCS I | 721 780 | 0.1891 | 0.5883 |
| BCS II | 722 686 | 0.1857 | 0.5605 |

Taking the BCS I as an example, the considered outage AC lines are listed in Table VII, and the optimization results are shown in Table VIII.

TABLE VII
CONSIDERED OUTAGE AC LINES

| Lines | Lines | Lines | Lines |
|---|---|---|---|
| L264(191-192) | L365(153-183) | L333(3-4) | L403(7039-39) |
| L114(59-61) | L396(7130-130) | L356(130-131) | L328(247-248) |
| L183(125-126) | L401(7012-12) | L344(45-46) | L340(21-20) |
| L88(39-42) | L354(121-115) | L345(62-61) | L338(15-17) |
| L346(63-64) | L358(132-170) | L341(24-23) | L320(242-245) |
| L305(225-191) | L347(73-74) | L324(244-246) | L322(243-244) |
| L366(155-156) | L353(116-124) | L327(246-247) | L335(7-6) |
| L290(214-215) | L349(85-99) | L319(240-281) | L330(249-250) |
| L177(119-120) | L348(81-88) | L339(16-15) | L337(12-10) |
| L399(7049-49) | L355(122-157) | L326(245-247) | L331(3-1) |
| L178(119-121) | L352(114-207) | L343(45-44) | L336(10-11) |
| L1(37-9001) | L357(130-150) | L321(242-247) | L332(3-2) |
| L402(7017-17) | L359(141-174) | L325(245-246) | L334(7-5) |
| L360(142-175) | L351(87-94) | L323(243-245) | L329(248-249) |
| L400(7139-139) | L350(86-102) | L342(36-35) | L173(118-119) |

TABLE VIII
OBJECTIVE FUNCTION VALUES BEFORE AND AFTER OPTIMIZATION

| Optimization status | $f_1$/($/h) | $f_2$/(p.u.) |
|---|---|---|
| Before optimization | 1 172 159 | 0.1974 |
| After optimization | 721 780 | 0.1891 |

Table VIII suggests that the generation cost $f_1$ and the voltage deviation index $f_2$ are respectively reduced by 38.42% and 4.20% by using the proposed optimization methodology. The above results suggest that the proposed approach manages to address the SC-MOPF problem of this system, and thereby its applicability to larger power systems is verified.

## VI. CONCLUSION

To balance economy and voltage quality, a SC-MOPF model is presented for a meshed AC/VSC-MTDC system, together with a Lasso-based contingency filtering scheme. Moreover, a solution approach based on PBCE-IBEA is developed to seek well-spread Pareto-optimal solutions via parallel computing, and thereby the integrated decision making is utilized to identify the BCSs. Studies performed on IEEE test systems reveal that our approach is capable of effectively achieving the trade-off between the economy and security for the meshed AC/DC system, and furthermore that the required computational time can be significantly shortened.

Future research will focus on considering dynamic indexes, such as the maximum of transient stability margin, as optimization objectives to cope with the dynamic security problems of the system. Besides, it is another interesting topic to investigate the MOPF for distribution systems with consideration of renewable generation and load uncertainties. What's more, considering the information of network is confidential when the VSC-HVDC link employed in the power exchanges between islands, the method for solving OPF problem with the data of DC networks remaining unknown is another interesting research study.

**REFERENCES**
[1] J. Luo, L. Shi, Y. Ni, "A solution of optimal power flow incorporating wind generation and power grid uncertainties," *IEEE Access*, vol. 6, pp. 19681-19690, Apr. 2018.
[2] B. Jiang, Z. Wang, and L. Kazmerski, "The key technologies of VSCMTDC and its application in China," *Renew. Sustain. Energy Rev.*, vol. 62, pp. 297–304, 2016.
[3] Y. Li, S. Wu, "Controlled islanding for a hybrid AC/DC grid with VSC-HVDC using semi-supervised spectral clustering," *IEEE Access*, vol. 7, no. 3, pp. 10478-10490, Jan. 2019.
[4] J. Momoh, R. Adapa, and M. El-Hawary, "A review of selected optimal power flow literature to 1993. Parts I and II," *IEEE Trans. Power Syst.*, vol. 14, no. 1, pp. 96–111, Feb. 1999.
[5] S. Duman. "A modified moth swarm algorithm based on an arithmetic crossover for constrained optimization and optimal power flow problems," *IEEE Access*, vol. 6, pp. 45394-45416.
[6] Y. Yang, A. Song, H. Liu, Z. Qin, J. Deng, and J. Qi, "Parallel computing of multi-contingency optimal power flow with transient stability constraints," *Protection Control Mod. Power Syst.*, vol. 3, no. 1, p. 20, 2018.

[7] T. Ding, C. Zhao, T. Chen, "Conic programming-based lagrangian relaxation method for DCOPF with transmission losses and its zero-gap sufficient condition," *IEEE Trans. Power Syst.*, vol. 32, no. 5, pp. 3852-3861, Dec. 2017.

[8] Y. Li, Z. Yang, G. Li, D. Zhao, W. Tian. Optimal scheduling of an isolated microgrid with battery storage considering load and renewable generation uncertainties. *IEEE Trans. Industr. Elec.* vol. 66, no. 2, pp. 1565-1575, Feb. 2019.

[9] Y. Li, Z. Yang, G. Li, Y. Mu, D. Zhao, C. Chen, B. Shen, "Optimal scheduling of isolated microgrid with an electric vehicle battery swapping station in multi-stakeholder scenarios: A bi-level programming approach via real-time pricing," *Applied Energy*, vol. 232, no. 2, pp. 603–608, Apr. 2002.

[10] C. Li, T. Ding, "An electric vehicle routing optimization model with hybrid plug-in and wireless charging systems," *IEEE Access*, vol. 6, pp. 27569-27578, May 2018.

[11] W. Zhang, Y. Xu, Z. Dong, K. P. Wong. Robust security constrained-optimal power flow using multiple microgrids for corrective control of power systems under uncertainty. *IEEE Trans. Industr. Inform.*, vol. 13, pp. 54-68.

[12] A. J. Monticelli, M. V. P. Pereira, and S. Granville, "Security-constrained optimal power flow with post-contingency corrective rescheduling," *IEEE Trans. Power Syst.*, vol. 2, no. 1, pp. 175–182, Feb. 1987.

[13] Y. Xu, Z. Y. Dong, R. Zhang, K. P. Wong, and M. Lai, "Solving preventive corrective SCOPF by a hybrid computational strategy," *IEEE Trans. Power Syst.*, vol. 29, no. 3, pp. 1345–1355, May 2014.

[14] K. Meng, W. Zhang, Y. J. Li, Z. Y. Dong, Z. Xu, K. P. Wong, and Y. Zheng, "Hierarchical SCOPF considering wind energy integration through multi-terminal VSC-HVDC grids", *IEEE Trans. Power Syst.*, vol. 32, no. 6, pp. 4211- 4221, Nov. 2017.

[15] W. Warid, H. Hizam, N. Mariun, and N. I. A. Wahab, "A novel quasi-oppositional modified Jaya algorithm for multi-objective optimal power flow solution," *Appl. Soft Comput.*, vol. 65, pp. 360-373, Apr. 2018.

[16] M.B. Shafik, Hongkun Chen, G.I. Rashed, and R. A. El-Sehiemy, "Adaptive multi objective parallel seeker optimization algorithm for incorporating TCSC devices into optimal power flow framework," *IEEE Access*, vol. 7, pp. 36934-36947, Mar. 2019.

[17] G. Chen, J. Qian, Z. Zhang, and Z. Sun, "Applications of novel hybrid bat algorithm with constrained Pareto fuzzy dominant rule on multi-objective optimal power flow problems", *IEEE Access*, vol. 7, pp. 52060-52084, Apr. 2019.

[18] X. Zhang, "Multiterminal voltage-sourced converter-based HVDC models for power flow analysis," *IEEE Trans. Power Syst.*, vol. 19, pp. 1877–1884, 2004.

[19] J. Beerten, S. Cole, and R. Belmans, "Generalized steady-state VSC MTDC model for sequential AC/DC power flow algorithms," *IEEE Trans. Power Syst.*, vol. 27, no. 2, pp. 821- 829, May 2012.

[20] J. Beerten, and R. Belmans, "Development of an open source power flow software for HVDC grids and hybrid AC/DC systems: MatACDC," *IET Gen., Transm., Distrib.*, vol. 9, no. 10, pp. 966-974, Jun. 2015.

[21] J. Cao, W. Du, H. F. Wang, and S. Q. Bu, "Minimization of transmission loss in meshed AC/DC grids with VSC-MTDC networks," *IEEE Trans. Power Syst.*, vol. 28, no. 3, pp. 3047-3055, Aug. 2013.

[22] M. Baradar, M. R. Hesamzadeh, and M. Ghandhari, "Second-order cone programming for optimal power flow in VSC-type AC-DC grids," *IEEE Trans. Power Syst.*, vol. 28, no. 4, pp. 4282-4291, Nov. 2013.

[23] W. Feng, A. Le Tuan, L. B. Tjernberg, A. Mannikoff, and A. Bergman, "A new approach for benefit evaluation of multiterminal VSC–HVDC using a proposed mixed AC/DC optimal power flow," *IEEE Trans. Power Del.*, vol. 29, no. 1, pp. 432-443, Feb. 2014.

[24] A. Rabiee, A. Soroudi, and A. Keane, "Information gap decision theory based OPF with HVDC connected wind farms," *IEEE Trans. Power Syst.*, vol. 30, no. 6, pp. 3396-3406, Nov. 2015.

[25] V. Saplamidis, R. Wiget, and G. Andersson, "Security constrained optimal power flow for mixed AC and multi-terminal HVDC grids," in *Proc. 2015 IEEE Eindhoven PowerTech*, 2015, pp. 1-6.

[26] J. Cao, W. Du, and H. F. Wang, "An improved corrective security constrained OPF for meshed AC/DC grids with multi-terminal VSC-HVDC," *IEEE Trans. Power Syst.*, vol. 31, no. 1, pp. 485-495, Jan. 2016.

[27] F. Sass, T. Sennewald, A. K. Marten, and D. Westermann, "Mixed AC high-voltage direct current benchmark test system for security constrained optimal power flow calculation," *IET Gen., Transm., Distrib.*, vol. 11, no. 2, pp. 447-455, 2017.

[28] Y. Li, B. Feng, G. Li, J. Qi, D. Zhao, and Y. Mu, "Optimal distributed generation planning in active distribution networks considering integration of energy storage," *Appl. Energy*, vol. 210, pp. 1073–1081, Jan. 2018.

[29] Y. Li, Z. Yang, D. Zhao, H. Lei, B. Cui, and S. Li, ''Incorporating energy storage and user experience in isolated microgrid dispatch using a multi-objective model,'' *IET Renew. Power Gener.*, vol. 13, no. 6, pp. 973–981, Apr. 2019.

[30] Y. Li, Y. Li, and G. Li, "A two-stage multi-objective optimal power flow algorithm for hybrid AC/DC grids with VSC-HVDC," In *Proc. 2017 IEEE PES General Meeting*, July 2017, pp. 1-5.

[31] Y. Li, Y. Li, G. Li, D. Zhao, and C. Chen, "Two-stage multi-objective OPF for AC/DC grids with VSC-HVDC: Incorporating decisions analysis into optimization process," *Energy*, vol. 147, pp. 286-296, Mar. 2018.

[32] M.K. Kim, "Multi-objective optimization operation with corrective control actions for meshed AC/DC grids including multi-terminal VSC-HVDC," *Int. J. Electr. Power Energy Syst.*, vol. 93, pp. 178-193, Dec. 2017.

[33] F. Capitanescu, M. Glavic, D. Ernst, and L. Wehenkel, "Contingency filtering techniques for preventive security-constrained optimal power flow," *IEEE Trans. Power Syst.*, vol. 22, no. 4, pp. 1690–1697, Nov. 2007.

[34] R. Sunitha, S. Kumar, and A. T. Mathew, "Online static security assessment module using artificial neural networks," *IEEE Trans. Power Syst.*, vol. 28, no. 4, pp.4328-4335, Nov.2013.

[35] Q. Jiang and K. Xu, "A novel iterative contingency filtering approach to corrective security-constrained optimal power flow," *IEEE Trans. Power Syst.*, vol. 29, no. 3, pp. 1099–1109, May 2014.

[36] W. Wang and M. Barnes, "Power flow algorithms for multi-terminal VSC-HVDC with droop control," *IEEE Trans. Power Syst.*, vol. 29, no. 4, pp. 1721–1730, Jul. 2014.

[37] F. Capitanescu and L. Wehenkel, "Improving the statement of the corrective security-constrained optimal power-flow problem," *IEEE Trans. Power Syst.*, vol. 22, no. 2, pp. 887–889, May 2007.

[38] R. Wiget, "Combined AC and multi-terminal HVDC grids—Optimal power flow formulations and dynamic control," Ph.D. dissertation, ETH Zurich, Zurich, Switzerland, 2015.

[39] M. Q. Li, S. X. Yang, and X. H. Liu, "Pareto or non-Pareto: bi-criterion evolution in multiobjective optimization," *IEEE Trans. Evol. Comput.*, vol. 20, no. 5, pp. 645-665, Oct. 2016.

[40] E. Zitzler, and S. Künzli, "Indicator-based selection in multiobjective search," in *Proc. 8th Int. Conf. Parallel Problem Solving Nature*, 2004, pp. 832- 842.

[41] Y. Kim, J. Hao, T. Mallavarapu, J. Park, and M. Kang, "Hi-LASSO: high-dimensional LASSO," *IEEE Access*, vol. 7, pp. 44562-44573, Apr. 2019.